\documentclass[12pt]{article}
\usepackage{amsmath,amssymb}

\newcommand{\be}{\begin{equation}}
\newcommand{\ee}{\end{equation}}
\newcommand{\ea}{\end{array}}
\newcommand{\beqa}{\begin{eqnarray}}
\newcommand{\eeqa}{\end{eqnarray}}

\newcommand{\nn}{\nonumber}
\def\tr{\mathop{\rm Tr}\nolimits}

\def\CP2{{\mathbb C}P^2}
\def\cliff{{{\mathbb C}\ell}}  

\def\half{\frac{1}{2}}
\def\I{{\mathbb I}}
\def\CDalign#1{\bgroup\vcenter\bgroup\tabskip 2pt 
       \baselineskip 14pt \lineskip 3pt \lineskiplimit 3pt
       \halign\bgroup &\hfill$##$\hfill\crcr
       #1\crcr\egroup\egroup\egroup} 

\newcommand{\gapproxeq}{\lower .7ex\hbox{$\;\stackrel{\textstyle
>}{\sim}\;$}}
\newcommand{\lapproxeq}{\lower .7ex\hbox{$\;\stackrel{\textstyle
<}{\sim}\;$}}

\newcounter{appendice}

\def\thebibliography#1{{\bf REFERENCES\markboth
 {REFERENCES}{REFERENCES}}\list
 {[\arabic{enumi}]}{\settowidth\labelwidth{[#1]}\leftmargin\labelwidth
 \advance\leftmargin\labelsep
 \usecounter{enumi}}
 \def\newblock{\hskip .11em plus .33em minus -.07em}
 \sloppy
 \sfcode`\.=1000\relax}

\begin{document}
\begin{titlepage}
\title{{\small\hfill SU-4252-749, DFUP-02-12}\\
Dirac Operators on Coset Spaces  }
\author{
A.P. Balachandran$^a$, Giorgio Immirzi$^b$, Joohan Lee$^c$,
 Peter Pre\v snajder$^d$\\
{\small\it $^a$ Physics Department,
Syracuse University}\\
{\small\it Syracuse NY 13244, USA}\\
{\small\it $^b$ Dipartimento di Fisica, Universit\`a di Perugia 
{\small\rm and} INFN, Sezione di Perugia,}\\
{\small\it Perugia, Italy}\\
{\small\it $^c$ Physics Department, University of Seoul, Seoul
130-743, Korea}\\
{\small\it $^d$ Department of Theoretical Physics, Comenius University}\\
{\small\it Mlynsk\'a dolina, SK-84248 Bratislava, Slovakia}
}
\maketitle
\begin{abstract}

The Dirac operator for a manifold $Q$, and its chirality operator when $Q$
is even dimensional, have a central role in noncommutative geometry.
We systematically develop the theory of this operator when $Q=G/H$, where
$G$ and $H$ are compact connected Lie groups and $G$ is simple.
 An elementary discussion
of the differential geometric and bundle theoretic aspects of $G/H$, including
its projective modules and complex, K\"ahler and Riemannian structures,
is presented for this purpose. An attractive feature of our approach
is that it transparently shows obstructions to spin- and spin$_c$-structures.
When a manifold is spin$_c$ and not spin, $U(1)$ gauge fields have to be
introduced in a particular way to define spinors 
\cite{{parthasarathy},{cahen}}.
Likewise, for manifolds like $SU(3)/SO(3)$, which are not even spin$_c$,
we show that $SU(2)$ and higher rank gauge fields have to be introduced
to define spinors. This result has potential consequences for string theories 
if such manifolds occur as $D$-branes. The spectra and eigenstates of the
Dirac operator on spheres $S^n=SO(n+1)/SO(n)$, invariant under $SO(n+1)$,
are explicitly found. Aspects of our work overlap with the earlier
research of Cahen et al. \cite{cahen}.

\end{abstract}
\end{titlepage}

\section{Introduction}

When a group $G$ acts transitively on a manifold $Q$ with stability group $H$
at a point $p$, we can identify $Q$ with the coset space $G/H$. Such spaces
are important in the description of Goldstone modes created by the spontaneous
breakdown of $G$ to $H$. Models of spacetime such as the Minkowski spacetime
$M^{3,1}$ or its compact Euclidean version $S^4$ are also of this sort.
The group $G$
in these cases is the Poincar\`e group and $SO(5)$ respectively, while $H$
is the Lorentz group and $SO(4)$ respectively. In addition, coset spaces like
${\mathbb C}P^N$ and $S^N$ have begun to proliferate as $D-$branes in string
and boundary conformal field theories.

The Dirac operator for a manifold $Q$, and its chirality operator when $Q$
is even--dimensional, have a central role in noncommutative geometry. That
is a good motivation for their study. This work focuses on this enterprise
when $Q$ is a coset space.
 In addition, in a subsequent paper, we shall develop fuzzy versions
of certain coset spaces and their Dirac and chirality operators, primarily
as a device to regularize quantum field theories thereon, and what
we do here is also a preparation for it.

We assume throughout that $G$ is a simple compact connected Lie group
and $H$ is a compact connected group. Without loss of generality we assume
also that $G$ is simply connected. These restrictions on $G$ and $H$ can
be relaxed somewhat, $G$ can be semi--simple for instance, and certain
noncompact Lie groups $G$ too seem approachable by our methods.

Not all $G/H$ admit a spin--, or even a spin$_c$--structure \cite{LM}.
One attractive aspect of our approach is that obstructions
to spin-- and spin$_c$--structure show up transparently and we can also
easily see when and how we can overcome them using suitable generalized
spin--structures, spin$_K$. The latter in general involve groups $K$ of any
dimension, whereas spin$_c$ (= spin$_{U(1)}$ in our notation) uses
$U(1)$ of dimension 1. The role of $K$ is roughly that of a gauge group,
so insisting on the existence of spinors introduces nontrivial gauge symmetry
and internal degrees of freedom. In addition, typically, spin$_K$-- theories
are chiral, left- and right--chiral spinors transforming differently
under $K$.  This suggests that there may be a clever way to use fuzzy spaces
to get the chiral fermions of the standard model.

There is a simple global appoach to differential geometry on $G/H$. We
introduce this formalism after setting up the preliminaries in Section 2.
We follow this up in Section 3 introducing spin-- and spin$_K$--structures.
Their Dirac and chirality operators are formulated in Section 4.
We call this version of the Dirac operator 'K\"ahler--Dirac operator', as
it is similar to the operator with the same name on a complex manifold.
There is another equivalent version using projective modules equally
useful for fuzzy physics, which we have decided to call the projective
Dirac operator. The `Dirac' operator then refers to either of these two
versions. Section 5 takes this up and also establishes its equivalence
to the K\"ahler--Dirac operator. Along the way, the differential geometry
of Section 2 is also translated to the language of projective modules.
The cut--off versions
of these expressions have an important role in fuzzy physics.
In Section 5 we also explicitly consider the spheres $S^{n}$
and  ${\mathbb C}P^n$. In particular, for spheres,we compute the
curvature and Dirac spectrum for the maximally symmetric metric.
Section 6 extends the preceding considerations to gravity on $G/H$ 
\cite{nair}. Finally Section 7 discusses the complex and K\"ahler 
structures of coset manifolds.

\section{Differential Geometry on $G/H$}
\setcounter{equation}{0}
\subsection{ Preliminaries}

$G$ is a simple, simply connected, connected, compact Lie group with
Lie algebra $\underline G$. $\underline H$ is a subalgebra of
$\underline G$ which by exponentiation generates a compact connected Lie group
$H$.

We think of $G$ concretely as $N\times N$ unitary matrices. The Lie algebra
$\underline H$ then has a basis $\{T(\alpha)\}$ of hermitean matrices
(we follow
physics conventions, more correctly $\{iT(\alpha )\}$ span  $\underline H$),
which are trace orthogonal:
\be
\tr T(\alpha)T(\beta)= c\,\delta_{\alpha\beta}\quad,\quad c=constant>0 \ .
\label{di}\ee
Using trace to define scalar product, $\underline G$ can be decomposed as
the orthogonal direct sum
\be
\underline G = \underline H \oplus^\perp \underline{G/H}\ .
\label{dii}
\ee
Let $\{S(i)\}$ be a basis for $\underline{G/H}$ with
\be
\tr S(i)S(j)=c\,\delta_{ij}\ .
\label{diii}
\ee
We also of course have
\be
\tr S(i)T(\alpha)=0\ .
\label{div}\ee
We denote the elements of the basis $\{ T(\alpha),S(i)\}$ collectively
as $\Sigma_A,\ A\in\{\alpha,i\}$.

Let $Ad$ denote the adjoint representation of $G$. Then $H\subset G$
leaves $\underline{G/H}$ invariant in this representation:
\be
h\,S(i)\,h^{-1}=S(j)(Ad\;h)_{ji}\quad,\quad h\in H \ .
\label{dv}\ee
We call this representation of $H$ on $\underline{G/H}$ as
$Ad_{G/H}$, and the corresponding representation of $\underline H$
as $ad_{G/H}$. $Ad_{G/H}(h)$ are real matrices
as the hermitean conjugation of (\ref{dv}) shows. They are also orthogonal as
conjugation leaves the relation (\ref{diii}) invariant. Thus if $|G|,\ |H|$ and
$|G/H|=|G|-|H|$ denote the dimensions of $G,\ H$ and $G/H$,
$\{ Ad_{G/H}(h)\}$ is a subgroup of $SO(|G/H|)$:
\be
\{Ad_{G/H}(h)\}\subseteq SO(|G/H|)\ .
\label{dvi}
\ee

The above discussion implies the following commutation  relations:
\beqa
\left[ T(\alpha ),T(\beta )\right]&=&ic_{\alpha\beta\gamma}T(\gamma)\ ,
\nonumber\\
\left[ T(\alpha ),S(i)\right] &=&ic_{\alpha ij}S(j)\ ,\label{dviia}\\
\left[ S(i),S(j)\right] &=&ic_{ij\alpha}T(\alpha )+ic_{ijk}S(k)\ .
\label{dvib}
\eeqa
The structure constants $c_{ABC}$ are real and totally antisymmetric.

We will call $c_{ijk}$ the torsion of the space $\underline{G/H}$. Below
we will see that it plays exactly the role
of the usual torsion for the canonical covariant
derivative on $G/H$ \cite{salam}. If $c_{ijk}=0$, the homogeneous space
$G/H$ is said to be `symmetric' \cite{helgason}. 
In that case, $\underline G$ admits the involutive automorphism:
\be
\sigma\; :\quad T(\alpha)\rightarrow T(\alpha)\ ,\quad S(i)\rightarrow
-S(i)
\label{dviii}
\ee
leaving $\underline H$ fixed. $\sigma$ lifts to an involutive automorphism
$\Sigma$ of $G$ leaving $H$ fixed, $\Sigma$ being defined from
\be
\Sigma\; :\quad  e^{i\theta_\alpha T(\alpha)}\rightarrow
e^{i\theta_\alpha T(\alpha)}\ ,\quad
 e^{i\theta_i S(i)}\rightarrow e^{-i\theta_i S(i)}\ .
\label{dix}\ee

\subsection{Tensor Fields on $G/H$ }

Let $W$ be a fixed vector space with an orthonormal basis $\{e_i\}$
which carries the representation $Ad_{G/H}$ of $H,\ h\;:\ e_i\rightarrow
e_j Ad_{G/H}(h)_{ji}$. The vector space $W^{\otimes n}=W\otimes W\otimes
\ldots\otimes W$ ($n$ factors) carries the tensor product representation
$Ad_{G/H}^{\otimes n}=Ad_{G/H}\otimes Ad_{G/H}\otimes\ldots\otimes
Ad_{G/H}$
($n$ factors). Let ${\mathbb C}\equiv W^{\otimes 0}$ also denote the
one--dimensional complex vector space carrying the trivial representation
$Ad_{G/H}^{\otimes 0}:\ h\rightarrow 1$.

Tensor fields of rank $n$ on $G/H$ can be defined to be equivariant functions
on $G$ with values in $W^{\otimes n}$. That means the following: for $n=0$
we have scalar fields $f^{(0)}$, complex (or $W^{\otimes 0}$) valued
functions on $G$ invariant under the right--action of $H$ on $G$
(equivariance):
\be
f^{(0)}= \hbox{scalar fields}:\quad f^{(0)}(gh)=f^{(0)}(g)\quad ,
\ \forall\; h\in H  \ .
\label{dx}\ee
A tensor field $f^{(1)}$ of rank 1 has values in $W$; we can write it as:
\be
f^{(1)}= f^{(1)}_i e_i, \quad f^{(1)}_i\ :\
g\rightarrow f^{(1)}_i(g)\in {\mathbb C}\ .
\label{dxi}
\ee
Equivariance for $n=1$ means the following transformation property under
the right action of $H$ on $G$:
\be
f^{(1)}_i(gh)e_i=f^{(1)}_i(g)Ad_{G/H}(h)_{ij}e_j\ .
\label{dxii}
\ee
Therefore
\be
f^{(1)}_i(gh)=f^{(1)}_j(g)Ad_{G/H}(h)_{ji}\ .
\ee

Let $J$ label the inequivalent irreducible representations of $G$ by unitary
 matrices $\{D^J(g)\}$; their matrix elements in a convenient orthonormal
basis are $D^J_{mn}(g)$. We have that
\be
D^J_{mn}(gh)=D^J_{mn'}(g)D^J_{n'n}(h)\ .
\label{dxiii}
\ee
 If the representation $h\rightarrow D^J(h)$ contains the identity
representation of $H$, we can choose the basis in the representation space
so that the index $n$  in (\ref{dxiii}) transforms trivially when $n\in$ an
appropriate index set $I_0$:
\be
D^J_{mi_0}(gh)=D^J_{mi_0}(g)\quad ,\ \forall\ i_0\in I_0\ .
\label{dxiv}
\ee
From this, (\ref{dx}) and Peter--Weyl theorem it follows that we can expand
$f^{(0)}$ in the form
\be
f^{(0)}(g)=\sum \xi^J_{m i_0} D^J_{mi_0}(g)\quad ,\quad\xi^J_{m i_0} \in
{\mathbb C}\ .
\label{dxv}\ee
$\xi^J_{m i_0} $ is zero if $h\rightarrow D^J(h)$ fails to contain the
trivial representation of $H$.

Henceforth we assume for notational simplicity that the
identity representation occurs only once in the restriction of the 
irreducible representations $J$ of $G$ to $H$, and so drop the 
index $i_0$ from $\xi^J_{m i_0}$.
Otherwise a degeneracy index has to be included  here and elsewhere.

In the same way, if the representation  $h\rightarrow D^J(h)$ contains
$Ad_{G/H}$, we can choose the basis in the representation space so that
the index $n$ in (\ref{dxiii}) transform by $Ad_{G/H}$ if $i,j$ belong 
to an appropriate index set $I$:
\be
D^J_{mi}(gh)= D^J_{mj}(g)[Ad_{G/H}(h)]_{ji}\ ,\quad i,j\in I\ .
\label{dxva}
\ee
(For notational simplicity we are assuming that $Ad_{G/H}$ occurs only once
in the representation $J$, otherwise a degeneracy index has to be added here
and elsewhere.)
Then we can expand $f^{(1)}_i$ in the form
\be
 f^{(1)}_i(g)=\sum \xi^J_m D^J_{mi}(g)\ ,\quad \xi^J_m \in {\mathbb C}\ .
 \ee
 $\xi^J_m $ now is zero if $h\to D^J(h)$ fails to contain $Ad_{G/H}$.

 Continuing in this vein we see that tensor fields of rank $n$ in component
form look like $f^{(n)}_{i_1...i_n}$ and have the expansion
 \be
 f^{(n)}_{i_1...i_n}(g)=\sum \xi^J_m D^J_{m,\{i_1...i_n\}}
 (g)\ ,\quad i_k\in I\ ,
 \ee
 \be
 D^J_{m,\{i_1...i_n\}}(gh)=D^J_{m,\{j_1...j_n\}}(g)[Ad_{G/H}(h)]_{j_1i_1}...
  [Ad_{G/H}(h)]_{j_ni_n}\ .
  \ee
We have used a convenient multi--index notation for the second index of
$D^J$. The rest should be clear. Tensor fields of diverse permutation
symmetries are readily constructed along similar lines.

\subsection{Covariant Derivative.}
Let ${\cal T}^{(n)}$ denote the space of tensor fields of rank $n$,
with a typical member $f^{(n)}=\{f^{(n)}_{i_1\ldots i_n}\}$.
 ${\cal T}^{(0)}$ consists of functions, and it is also an algebra under
pointwise multiplication. All  ${\cal T}^{(n)}$ are
${\cal T}^{(0)}$-modules. The covariant derivative $\nabla$ is a map
\be
\nabla\ :\quad  {\cal T}^{(n)}\ \rightarrow\  {\cal T}^{(n+1)}\quad ,
\quad f^{(n)}\ \rightarrow\ \nabla f^{(n)}\ ,
\label{dxvi}
\ee
where $\nabla f^{(n)}$ has components $(\nabla f^{(n)})_{ii_1\ldots i_n}$.
It has in addition to fulfill the following important derivation
property.
Note that we can take tensor products of ${\cal T}^{(n)}$-s (over
${\cal T}^{(0)},\ {\cal T}^{(n)}$ being  ${\cal T}^{(0)}$ modules):
\be
{\cal T}^{(n)}\otimes{\cal T}^{(m)}={\cal T}^{(n+m)}\quad ,\quad
f^{(n)}\otimes f^{(m)}=f^{(n+m)}\ ,
\label{dxvii}\ee
where
\be
f^{(n+m)}_{i_1\ldots i_nj_1\ldots j_m}=f^{(n)}_{i_1\ldots i_n}
f^{(m)}_{j_1\ldots j_m}\ .
\ee
 Then we require that
\be
\nabla(f^{(n)}\otimes f^{(m)})=\nabla f^{(n)}\otimes f^{(m)}+
f^{(n)}\otimes \nabla f^{(m)}\ .
\label{dxviii}
\ee

There is a natural choice for the covariant derivative in our case.
We call it hereafter as $X$. The action of $X$ on functions is:
\be
\left[ Xf^{(0)}\right]_i(g)=\frac{d}{dt}\;f^{(0)}(ge^{itS(i)})\Big|_{t=0}\ .
\label{dxix}\ee
$\nabla f^{(0)}$ transforms correctly in view of (\ref{dv}).
In the same way the action on $f^{(n)}$ is:
\be
\left[ Xf^{(n)}\right]_{i i_1\ldots i_n}(g)=
\frac{d}{dt}\;f^{(n)}_{i_1\ldots i_n}(ge^{itS(i)})\Big|_{t=0}\ .
\label{dxx}\ee
The right--hand side defines a vector field $X_i$. Using $X_i$ the
covariant derivative in components is
$f^{(n)}_{i_1\ldots i_n}\ \rightarrow\ X_if^{(n)}_{i_1\ldots i_n}=$
R.H.S. of (\ref{dxx}).

The torsion of the covariant derivative vanishes only if
$\left[ X_i,X_j  \right] f^{(0)}=0$. From the definition and
(\ref{dviia}), we have
\be
\left[ X_i,X_j  \right] f^{(0)}=-c_{ijk}X_kf^{(0)}\ .
\label{dxxi}\ee
So there is torsion if $c_{ijk}\ne 0$. But there is an easy way to construct
the torsion--free covariant derivative $\overline X_i$. Set
\beqa
\overline X_i f^{(0)}&=&X_if^{(0)}\;,\nonumber \\
\overline X_i f^{(1)}_j&=&X_if^{(1)}_j+\frac{1}{2}c_{ijk}f^{(1)}_k\ ,
\label{dxxii} \\             &\ldots&  \nonumber\\
\overline X_i f^{(n)}_{j_1\ldots j_n}&=&X_if^{(n)}_{j_1\ldots j_n}+
\frac{1}{2}c_{ij_1j_1'}f^{(n)}_{j_1'j_2\ldots j_n}+
\frac{1}{2}c_{ij_2j_2'}f^{(n)}_{j_1j_2'\ldots j_n}+\ldots +
\frac{1}{2}c_{ij_nj_n'}f^{(n)}_{j_1\ldots j_n'}\ .
\nonumber
\eeqa
Then
\be
\left[ \overline X_i,\overline X_j  \right] f^{(0)}=
\left[  X_i, X_j  \right] f^{(0)}
+c_{ijk}X_kf^{(0)}=0\ ,
\label{dxxiii}\ee
just as we want.

Gauge fields will certainly have a central role in further developments.
So we briefly indicate what they are here. Let us first consider $U(1)$
gauge fields. The general gauge potential is $A_i=\sum\xi^J_M D^J_{Mi}\ $,
$\xi^J_M\in \mathbb C$. It is subject to the reality condition
$\overline A_i=-A_i$. Then if $f^{(n)}$ has charge $e$, its covariant
derivative is $(\overline X_i+eA_i)f^{(n)}_{i_1\ldots i_n}$, where $A_i$
acts by pointwise
multiplication. This definition is compatible with equivariance.
We can substitute $X_i$ for $\overline X_i$ at the cost of possible torsion.

The gauge covariant derivative for a general gauge group as usual only
involves regarding $eA_i(g)$, that is $e\xi^J_M$, to be Lie algebra
valued, its action on $f^{(n)}$ in (\ref{dxx}) is then dictated by the
representation content of the latter.

\section{Spin-- and Spin$_K$--structures}
\setcounter{equation}{0}

Spinorial fields are essential for physics. We can go about constructing
them as follows. The orthogonal group $SO(|G/H|)$ has a double cover
$Spin(|G/H|)$. Associated with $SO(|G/H|)$, there is also a Clifford algebra
$\cliff(|G/H|)$ with generators $\gamma_1,\gamma_2,\ldots ,\gamma_{|G/H|}$:
\be
\gamma_i\gamma_j+\gamma_j\gamma_i=2\delta_{ij}\, \I \ .
\label{dxxiv}\ee
Here $\I$ denotes unit matrix. (Its dimension should be clear from the 
context). $\cliff(|G/H|)$ has one or two inequivalent IRR's
of dimension $2^n$\ \ if $|G/H|=2n$ or $|G/H|=2n+1$.
In the latter case, the two IRR's are related by
a change of sign of all $\gamma_i$'s. In either case, they generate a
unique faithful representation of $Spin(|G/H|)$ with generators 
$\Sigma_{ij}=\frac{1}{4i}(\gamma_i\gamma_j-\gamma_j\gamma_i)$ which we call
$Spin^\cliff(|G/H|)$.

A recursive scheme for constructing anticommuting sets of hermitean 
$\gamma-$matrices goes as follows. We start with a set of 
$2^{n-1}\times 2^{n-1}$ matrices $\gamma_i,\ i=1,...,2n-1$, satisfying 
eq.(\ref{dxxiv}), and such that $(-i)^{n-1}\gamma_1...\gamma_{2n-1}=1$,
e.g. for $n=2$, the three Pauli matrices. Then a set
of $2^n\times 2^n$ matrices $\Gamma_\lambda,\ \lambda=1,...,2n+1$, 
satisfying eq.(\ref{dxxiv}), and such that 
$(-i)^n\Gamma_1...\Gamma_{2n+1}=\I$ is given by
\be
\Gamma_i=\left(\begin{matrix}0&\gamma_i\\ \gamma_i&0\end{matrix}\right),
\ i=1,...,2n-1,\quad
\Gamma_{2n}=\left(\begin{matrix}0&-i\I\\ i\I &0\end{matrix}\right),\quad 
\Gamma_{2n+1}=\left(\begin{matrix}\I &0\\ 0&-\I \end{matrix}\right)\ .
\label{dxxiva}
\ee
The matrices $\gamma_1,...,\gamma_{2n-1}$ span $\cliff(2n-1)$, and the 
matrices $\Gamma_1,...,\Gamma_{2n}$ span $\cliff(2n)$, whereas
$\Gamma_1,...,\Gamma_{2n+1}$ span $\cliff(2n+1)$.

\subsection{ Spin Manifolds}

{\it We say that $G/H$ is a spin manifold if the commutative diagram of Fig.1
exists, arrows being homomorphisms (which need not be onto):}
\be
\begin{matrix}
    G&\supset& H         &\longrightarrow& Spin^\cliff(|G/H|)\\
     &       & \downarrow&              &\downarrow^{{\mathbb Z}_2}\\
     &       &Ad_{G/H}   & \subset      &SO(|G/H|)\\
     &       &           &              &         \\
     &       &           &\hbox{Fig. 1} &
\end{matrix}\nonumber
\ee
The vertical homomorphisms are there by construction, so what is to be
verified is the existence of the horizontal arrow.
If it exists, a general spinor can be constructed as follows. We can reduce
$Spin^\cliff(|G/H|)$ restricted to $H$ into a direct sum $\oplus\rho$ of
unitary
irreducible representations of $H$. Let $g\rightarrow D^J(g)$ be the unitary
matrix of $g$ in a representation of $G$ which on restriction to $H$
contains $\oplus\rho$. Then we can restrict its second index $a$
to an index set
$I$ so that it transforms by $\oplus\rho$ under $g\rightarrow gh$:
\be
D^J_{\alpha a}(gh)=D^J_{\alpha b}(g)D^J_{ba}(h)\quad ,\quad a,b\in I\ .
\label{dxxv}
\ee
By construction we know how the Clifford algebra acts on the index $a\in I$.
A general spinor $\psi$ then is a function on $G$ with components
\be
\psi_a=\sum\xi_M^J D^J_{M a}\ .
\label{dxxvi}\ee
Let us look at examples.

\medskip
\noindent Example 1: ${\mathbb C}P^1=SO(3)/SO(2)=
[Spin(3)=SU(2)]/[Spin(2)=U(1)]$.
 So $G=SU(2),\ H=U(1)=\{ e^{i\sigma_3\theta/2}\}$, $\sigma_A$
the Pauli matrices. Then $S(i)=\sigma_i,\ i=1,2$, and
\be
e^{i\sigma_3\theta/2}\sigma_ie^{-i\sigma_3\theta/2}=\sigma_j
R_{ji}(\theta)\ ,\quad R(\theta)=\left(\begin{matrix}
\cos\theta&\sin\theta\cr -\sin\theta&\cos\theta\cr \end{matrix}
\right)\ \in\ SO(2)\ .
\label{dxxvii}\ee
$Spin^\cliff(|G/H|)$ is just $H=\{e^{i\sigma_3\theta/2}\}$, the homomorphism
$Spin^\cliff(2)\rightarrow SO(2)$ being
$e^{i\sigma_3\theta/2}\rightarrow R(\theta)$.
Thus Fig.1 exists and $SU(2)/U(1)=S^2\simeq {\mathbb C}P^1$ is spin.

For a thorough treatment of noncommutative geometry and Dirac operator
on $S^2$, see \cite{mignaco}. 

\noindent Example 2: Similar arguments show that all the spheres
$S^N=SO(N+1)/SO(N)=Spin(N+1)/Spin(N)$ are spin. $G$ for $S^N$ is $Spin(N+1)$
while $H=Spin(N)$. $Ad_{G/H}$ is $SO(N)$, the ${\mathbb Z}_2$--quotient of
$Spin(N)$. Since $Spin^\cliff(|G/H|)$ is isomorphic to $Spin(N)$, $S^N$
is spin.

\noindent Example 3: ${\mathbb C}P^2=SU(3)/U(2)$. So $G=SU(3),\ H=U(2)$.
A basis for the 3--dimensional $SU(3)$-- Lie algebra consists of the
Gell-Mann matrices $\lambda_A$. The $U(2)$ Lie algebra  has basis
$\lambda_1,\lambda_2,\lambda_3,\lambda_8$, the hypercharge $Y$ being
$\frac{1}{\sqrt 3}\lambda_8$. The $S(i)$ are $\lambda_4,\lambda_5,
\lambda_6,\lambda_7$. Under $U(2)$, they transform as $(K^+,K^0)$ or
$(-\overline K^0,K^-)$ in particle physics notation. That means that
$Ad_{G/H}=U(2)$.  Regarding $U(2)$ as $2\times 2$ unitary matrices $U$,
we can embed $U(2)$ in $SO(4)$ by the map
\be
U\rightarrow \frac{1}{2}\left(
\begin{matrix} U+U^*& i(U-U^*)\cr -i(U-U^*)&U+U^*\cr\end{matrix}\right) \ .
\label{dxxviii}
\ee
$Spin^\cliff(4)$ is the $(\frac{1}{2},0)\oplus (0,\frac{1}{2})$ representation
of $SU(2)\otimes SU(2)$. It is the double cover of $SO(4)$.
Now $H$ is $U(2)$ and $Spin^\cliff(4)$ has no $U(2)$ subgroup.
So ${\mathbb C}P^2$ is not spin \cite{parthasarathy}.

\noindent Example 4: $G=SU(3),\ H=SO(3)$.
With $G$ as $3\times 3$ unitary matrices, $H$ consists of all real orthogonal
matrices and corresponds to the spin 1 representation of $SO(3)$.
$\underline{G/H}$ is of dimension 5. It carries the spin 2 representation
of $SO(3)$, isomorphic (but not equivalent!) to the spin 1 representation.
There is no homorphism $SO(3)\rightarrow Spin^\cliff(5)$ compatible 
with Fig.1, so that $SU(3)/SO(3)$ is not spin \cite{LM}.

 Let us show this result in more detail.
We can show it by establishing that the $2\pi$--rotation in $SO(3)$ becomes
a noncontractible loop in $SO(5)$ under the embedding in Fig.2. Then
the inverse image of $SO(3)$ under the homomorphism 
$Spin^\cliff(5)\rightarrow SO(5)$ is $SU(2)$ giving us the result.

Now $SO(3)$ acts on real symmetric traceless $3\times 3$ matrices 
$T=(T_{ij})$
according to $T\rightarrow RTR^T$. This is its spin 2 representation.
We can eliminate say $T_{33}$ using $\tr T=0$, thereby representing it as
real transformations on $(T_{11},T_{12},T_{13},T_{22},T_{23})$. $SO(5)$
consists of real transformations on this five-dimensional vector,so we now
have the needed explicit embedding of $SO(3)$ in $SO(5)$. Let
\be
R(\theta)\;:\quad\left(\begin{matrix}
\cos\theta & \sin\theta & 0\cr -\sin\theta & \cos\theta &0\cr 0&0&1
\end{matrix}\right)\ .
\label{dxxix}\ee
It generates the $2\pi$--rotation loop in $SO(3)$ as $\theta$ increases
from $0$ to $2\pi$.
Consider $T'=R(\theta)TR(\theta)^T$. Then as $\theta$ increases
from $0$ to $2\pi$ we have a $2\pi$--rotation in the $T_{13}\; -\; T_{23}$
plane. But by the time $\theta=\pi$, $T_{ab}\ (a,b\le 2)$ return to
$T_{ab}$ so that the rotations in their planes are by $4\pi$ or/and $0$
(in fact $\delta_{ab}(T_{11}+T_{22})$ undergoes no change and
$T_{ab}-\frac{1}{2}\delta_{ab}(T_{11}+T_{22})$ undergoes $4\pi$--rotation).
The corresponding loop of matrices in $SO(5)$ is the product of an odd
number of $2\pi$--rotations and hence cannot be deformed to a point 
in $SO(5)$. That concludes the proof.

We will now explain the Dirac operators for Spin-- and Spin$_K$--manifolds
after discussing Spin$_K$-structures.

\subsection{Spin$_K$ --Manifolds}

$K$ and $\cal H$ are compact connected Lie groups in what follows.
{\it We say that $G/H$
is a Spin$_K$--manifold if the commutative diagram of Fig.2 exists.
}
\be
\begin{matrix}
G&\supset&H&\longrightarrow&{\cal H}&\supset&Spin^\cliff(|G/H|)\\
 &       &\downarrow&  &\ \ \downarrow^K&  \swarrow{{\mathbb Z}_2}& \\
 &       &Ad_{G/H}  &        \subset    &SO(|G/H|)  &  &           \\
 &       & &        &                &              &              \\
 &       & &\hbox{Fig. 2}  & & &
\end{matrix}
\nonumber
\ee

$K$ and ${\mathbb Z}_2$ on the arrows are to show that they are the kernels
 of those homomorphisms.

$Spin_{U(1)}$ in our language is what mathematicians call $Spin_c$.

The intersection $Spin^\cliff(|G/H|)\cap K$ clearly contains 
${\mathbb Z}_2$.
It cannot be larger, for that would mean that the kernel for the slanting
arrow exceeds ${\mathbb Z}_2$.\\
 Thus ${\cal H} \supset
\left[ Spin^\cliff(|G/H|)\times K \right] /{\mathbb Z}_2 $.
Its quotient by $K$ being exactly $SO(|G/H|)$, we conclude that
\be
{\cal H}=\left[ Spin^\cliff(|G/H|)\times K \right] / {\mathbb Z}_2\ ,
\label{dxxx}
\ee
giving Fig.3, also a commutative diagram:
\be
\begin{matrix}
G&\supset&H&\longrightarrow &
  {\cal H}=\left[ Spin^\cliff(|G/H|)\times K \right] / {\mathbb Z}_2
                                            &\supset & Spin^\cliff(|G/H|)\\
 & & & & & & \\
 &  &\downarrow&  &\ \ \downarrow K&  \swarrow{\mathbb Z}_2\ \  &  \\
 & & & & & & \\
               &   &Ad_{G/H}&\subset&SO(|G/H|)  &   & \\
 & & & & & & \\
& & & & \hbox{Fig. 3} & &
\end{matrix}
\nonumber
\ee

Let us denote the generators of ${\mathbb Z}_2$ in $Spin^\cliff(|G/H|)$
and $K$ by
$ z_{Spin}$ and $z_K$, they square to the respective identities. The
inclusion of $Spin^\cliff(|G/H|)$ in Fig.3 is to be understood as follows.
The elements of ${\cal H}$ are the equivalence classes
\be
 <s,k>=<z_{spin}s,z_K k>\ ,\quad s\in Spin^\cliff(|G/H|)\ ,\quad k\in K \ .
\label{dxxxi}\ee
Then the top inclusion is via the isomorphism
\be
  s\ \rightarrow\ <s,e_K>\ ,\quad e_K=\ \hbox{identity of}\ K \ .
\label{dxxxii}
\ee

As we think of ${\cal H}$ as the concrete matrix group obtained by tensoring
$Spin^\cliff(|G/H|)$ with a faithful unitary representation of $K$
where $z_K$ is represented by $-\mathbb I$
\be
{\cal H}=Spin^\cliff(|G/H|)\otimes K
\label{dxxxiii}\ee
we can write $- 1$ for $z_{spin}$ and $z_K$. The inclusion of
$Spin^\cliff(|G/H|)$ is then just $s\;\rightarrow\;s\otimes \mathbb I$.

Let us motivate the new requirements in Figures 2 and 3. For a physicist, a
spinor changes sign under `$2\pi$--rotation'. ${\cal H}$ is the group
acting on $Spin_K$--spinors. We have required it to contain 
$Spin^\cliff(|G/H|)$,
so we can check this requirement by looking at the action of 
$2\pi$--rotation $\in Spin^\cliff(|G/H|)\subset{\cal H}$.
As for asking that $H\rightarrow{\cal H}$,
we can reduce the representation of ${\cal H}$ into a direct sum $\oplus\rho$
of irreducible representations $\rho$ of $H$ just as in the discussion 
of spin structures. The action of the Clifford algebra on 
$\oplus\rho$ by construction is known. 
The wave functions of $Spin_K$--spinors are then given by linear
spans of representations of $G$ induced by $\oplus \rho$, see (\ref{dxxvi}).
Later we shall see how the Dirac operator can be defined on these wave
functions.

\medskip
\noindent Example 5. $G=SU(3),\ H=U(2),\ G/H={\mathbb C}P^2$.
Here we choose $K=U(1)$. Elements of $U(2)$ can be written as the 
equivalence classes
\be
< s,u>=< -s,-u>\ ,\quad s\in SU(2)\ ,\quad u\in U(1)\ ,
\label{dxxxiv}\ee
where we identify $SU(2)$ with $2\times 2$ unitary matrices of unit
determinant and $u$ with a phase. $Spin^\cliff(4)$ is 
$SU(2)\otimes SU(2)$ and ${\cal H}$ consists of the equivalence classes
\be
<s_1,s_2,u>=<-s_1,-s_2,-u>\ .
\label{dxxxv}
\ee
With elements of $SO(4)$ represented as $<s_1,s_2>=<-s_1,-s_2>$, the
homomorphism ${\cal H}\rightarrow SO(4)$ is $<s_1,s_2,u>\rightarrow 
(s_1,s_2)$. The homomorphism $H\rightarrow{\cal H}$ is also simple:
\be
\left[ s,u\right]\ \rightarrow\ <s,s,u> \ .
\label{dxxxvi}
\ee
Thus ${\mathbb C}P^2$ is $Spin_{U(1)}$ or $Spin_c$.

\medskip
\noindent Example 6. $\ G=SU(3),\ H=SO(3)$.

We return to the choices $G=SU(3),\ H=SO(3)$. Choosing $K=U(1)$ is not
helpful now, as we lack a suitable homomorphism $H=SO(3)\rightarrow
{\cal H}=Spin^\cliff(5)\otimes_{ Z_2}U(1)$. So $SU(3)/SO(3)$ is not even
Spin$_{U(1)}$, a result originally due to Landweber and Stong \cite{LM}.
A better choice is $K=SU(2)$. Then
we can find the homomorphism $H\rightarrow{\cal H}$ as follows. The image
of $Ad_{G/H}$ in $SO(5)$ is an $SO(3)$ subgroup $SO(3)'$. Its inverse
image in $Spin^\cliff(5)$ is an $SU(2)$ subgroup $SU(2)'$. Let $\vec\Sigma$
and $\vec T$ be the angular momentum generators of $SU(2)'$ and $K$.
If $\vec L$ are the angular momentum generators of $H$, the map at the level
of Lie algebras is just $\vec L\rightarrow \vec\Sigma+\vec T$. Hence
$SU(3)/SO(3)$ is Spin$_{SU(2)}$. More such examples can be found.

\subsection {What is $\overline X (i)$ now?}

We need the extension of the torsion--free connection with components
$\overline X_i$ to spinors on general $spin_K$--manifolds.  The first step
in this direction is the extension of $X_i$.

A spinor field $\psi=(\psi_a)$ on a $Spin_K$--manifold has the expansion
\be
\psi_a=\sum \xi^L_M D^L_{Ma}\ ,
\label{dxxxvii}
\ee
where $a$ carries the action of the Clifford algebra. The definition of
$X_i$ on $\psi$ is immediate from (\ref{dxix}):
\be
(X_i\psi)(g)=\frac{d}{dt}\sum\xi^L_M D^L_{Ma}(ge^{itS(i)})\big|_{t=0}\ .
\label{dxxxviii}
\ee

The definition of $\overline X_i$ involves the extension of $c_{ijk}$ to
spinors so that it can act on the index $a$. Now, the generators of
the $SO(|G/H|)$-Lie algebra are $M_{ij}$ where:
\be
(M_{ij})_{kl}=-i(\delta_{ik}\delta_{jl}-\delta_{il}\delta_{jk}) \ .
\label{dxxxix}\ee
Its image in the spinor representation is $\frac{1}{4i}[\gamma_i,\gamma_j]$.

Now
\be
c_{ij'k'}(M_{j'k'})_{kj}f^{(1)}_k=-2i c_{ikj}f^{(1)}_k \ .
\label{dxl}\ee
Hence
\be
\overline X_if^{(1)}_j=X_if^{(1)}_j+\frac{1}{4i}c_{ij'k'}(M_{j'k'})_{kj}
f^{(1)}_k \ .
\label{dxli}
\ee
From this follows the definition of $\overline X_i$ on spinors:
\be
\overline X_i\psi_a= X_i\psi_a-\frac{1}{16}c_{ijk}
\left(\left[\gamma_{j},\gamma_{k}\right]\right)_{ba}\psi_b\ .
\label{dxlii}\ee
A tensor formed of spinors will then transform correctly.

The introduction of gauge fields follows the earlier discussion.

\section{The Dirac and Chirality Operators}
\setcounter{equation}{0}

The massless Dirac operator for the torsion free connection $\overline X_i$
is just:
\be
D_W= -i\gamma_i^{\cal R}\overline X_i\ \ ,
\label{dxliv}\ee
where the superscript $\cal R$ indicates that the $\gamma$'s act on
spinors on the right.
It is self-adjoint. This expression can be gauged, and $\overline X_i$
can be substituted by $X_i$ if we can tolerate torsion.

If $|G/H|$ is even, e.g. if $G/H$ is a [co]adjoint orbit,
there is also a chirality operator $\gamma$ anticommuting
with $D_W$:
\be
\gamma= (-i)^{\frac{1}{2}|G/H|}\;\gamma_1\ldots\gamma_{|G/H|}=
\gamma^\dagger\ ,\qquad \gamma^2=\I \ .
\label{dxlv}
\ee

The subscript `$W$' is to indicate that it is the form of the Dirac operator
used by the Watamuras \cite{W}. For even $|G/H|$
there is also the unitarily equivalent Dirac
operator \cite{peter}
\be
D=e^{i\gamma^{\cal R}\pi/4}D_We^{-i\gamma^{\cal R}\pi/4} =
i\gamma^{\cal R} D_W\ ,
\label{dxlvi}\ee
which is central to fuzzy physics.

\section{Projective Modules and their Dirac Operator}
\setcounter{equation}{0}

\subsection{Projective Modules.}

In the algebraic approach to vector bundles, their sections are substituted
by elements of projective modules (`of finite type') \cite{V}.
A projective module
is constructed as follows. Let $A$ be an algebra. It can be the
commutative algebra $\cal A$ of $C^\infty$--functions on a manifold $M$ if
our interest is in the algebraic description of its vector bundles. But it
can also be a noncommutative algebra, in which case there is no evident
correspondence with sections of differential geometric vector bundles.
Consider $A^N\equiv A\otimes_C{\mathbb C}^N$ with elements
$a=(a_1,\ldots,a_N),\ a_i\in A$. Let $P$ be an $N\times N$ projector with
coefficient in $A$:
\be P_{ij}\in A\ ,\quad P^\dagger=P=P^2\ .
\label{dxlvii}\ee
Then $A^NP$ (whose elements are vectors $\alpha$
with components $a_jP_{ji}$) is a projective module.
The Serre--Swan theorem (\cite{V}) establishes that sections of any 
vector bundle can be got from some $N$ and $P$.

It is very helpful for subsequent developments to have a projective module
description of vector bundles. We can find the appropriate projectors by a
known method described nicely by Landi \cite{GL}. It goes as follows.

Consider for example a rank 1 tensor field and any particular $D^J$ matrix
occurring in its expansion, with elements $D^J_{\rho i}$. We have
\be
(D^J)^\dagger_{i\rho}(g)D^J_{\rho j}(g)=\delta_{ij} \ .
\label{dxlviii}\ee
Let
\be
P^J(g)_{\rho\sigma}=D^J_{\rho i}(g)(D^J)^\dagger_{i\sigma}(g) \ .
\label{dxlix}\ee
Since $P^J(gh)=P^J(g)$ if $h\in H$, $P^J_{\rho\sigma}$ are functions on $G/H$.
In view of (\ref{dxlviii}), they are projectors too. If $P^J(g)$ are
$|J|\times |J|$ matrices, a projective module describing rank 1 tensor
fields is
\be
{\cal A}^{|J|}P^J=<\alpha^J=(\alpha^J_1,\ldots,\alpha^J_{|J|}),
\ \alpha^J_i=a^J_kP^J_{ki}> \ .
\label{dl}
\ee

There is no unique correspondence between projective modules and vector
bundles. Thus for each $J$, we can find a projector and its module. But all
such modules are equivalent, since there are elements $\alpha^J=a^JP^J$
and $\alpha^K=a^KP^K$ which naturally correspond for different $J$ and $K$:
\be
\alpha^J_\rho=\sum\xi^L_MD^L_{Mi}(D^J)^\dagger_{i\rho}\quad ,\quad
\alpha^K_\sigma=\sum\xi^L_MD^L_{Mi}(D^K)^\dagger_{i\sigma} \ .
\label{dli}
\ee

\subsection{Differential Geometry}
There is much to be said on the differential geometry on projective
modules, but for reasons of brevity we limit ourselves
to indicating how to extend the definitions of
$X(i)$ and $\overline X(i)$.

Let us first focus on the tensorial case. Let
\be
P=(P_{\lambda\rho})\ ,\quad P_{\lambda\rho}=
D^J_{\lambda i}(D^J)^\dagger_{i\rho}
\label{dlxxvii}\ee
be a projector appropriate for rank 1 tensors. Then what substitutes for the
torsion--free $X_i$ acting on $\alpha^J$ is $\nabla_\lambda$, which is 
defined by
\be
\nabla_\lambda\alpha^J_\rho(g)=\sum\xi^L_M\frac{d}{dt} D^L_{Mi}
(ge^{itS(j)})\Big|_{t=0} (D^J)^\dagger_{i\rho}(g)(D^J)^\dagger_{j\lambda}(g)
\ .\label{dlxxix}
\ee
It belongs to the projective module for rank 2 tensors with the natural
choice $P\otimes P$ of their projectors. A more compact expression for the
covariant derivative can be given in terms of the right-invariant vector
fields of $G$, defined by
\be
({\cal L}_Af)(g)=-i\frac{d}{dt}f(e^{-it\Sigma_A}g)\big|_{t=0}\ ,
\label{dlvii}
\ee
so that
\be
{\cal L}_A(D^J)^\dagger_{\lambda\mu}=(D^J)^\dagger_{\lambda\rho}
(\Sigma_A)_{\rho\mu}^J  \ .
\label{dlviii}
\ee
If functions on $G/H$ are regarded as functions on $G$ invariant by right
action of $H$ on $G$, ${\cal L}_A$ on these correspond to `orbital'
operators of angular momentum.

The vector fields  ${\cal L}_A$ are related to the left--invariant vector
fields $X_A$ by:
\be
(X_Af)(g)=\frac{d}{dt}f(ge^{it\Sigma_A})\big|_{t=0}=
-i({\cal L}_Bf)(g)\,(Ad\,g)_{BA}\ .
\label{dliii}
\ee
From this we may derive the following expression for the covariant 
derivative:
\be
\nabla_\lambda =-i\,Ad\,g_{Ai}\,(D^J)^\dagger_{i\lambda}{\cal J}_A\ ,\quad
{\cal J}_A=\ \hbox{`total angular momentum'}\ ={\cal L}_A-\Sigma_A^{\cal R}
\ ,\label{dlxxviii}
\ee
where $\Sigma_A^{\cal R}$ are the generators appropriate for representation $J$
acting on the right. In fact, applying (\ref{dliii}) and then
(\ref{dlviii}) to the definition (\ref{dlxxix}) we find
\be
\nabla_\lambda\alpha_\rho^J=-i\big({\cal L}_A\sum\xi^L_MD^L_{Mi}\big)
D^{J\dagger}_{i\rho}(Ad\,g)_{Aj}D^{J\dagger}_{j\lambda}=
-i(Ad\,g)_{Aj}D^{J\dagger}_{j\lambda}\Big(({\cal L}_A-\Sigma_A^{\cal R})
\alpha^J\Big)_\rho  \ .
\ee
$\nabla_\lambda$  maps
tensors of rank $k$ to $k+1$. It also has the correct derivation property so
that it is a covariant differentiation. Also (\ref{dlxxix}) shows that it
corresponds to the operator $X_i$.

We can define the covariant derivative $\nabla_\rho$ on spinors 
 corresponding to $X_i$ in the same way, just changing the index $i$ to $a$
in (\ref{dxxxvii}), and accordingly changing the choice of $J$ as well.

The canonical torsion $c_{ijk}$ generalizes for tensors to
\be
{\cal C}_{\lambda\rho\sigma}=(D^J)^\dagger_{i\lambda}(D^J)^\dagger_{j\rho}
c_{ijk}D^J_{\sigma k} \ .
\label{dlxxx}
\ee
A torsion--free covariant derivative on tensors when $c_{ijk}\ne 0$ is then
defined from
\be
\overline\nabla_\lambda\alpha^J_\rho=\nabla_\lambda\alpha^J_\rho+
\frac{1}{2}{\cal C}_{\lambda\rho\sigma}\alpha^J_\sigma \ .
\label{dlxxxiiib}
\ee

As for spinors, following (\ref{dxlii}), we define a spinorial torsion
which is twice the expression
\be
-\frac{i}{4}c_{ijk}\frac{1}{2i}(\gamma_{j}\gamma_{k})_{ba}=
-\frac{1}{8}c_{ijk}(\gamma_{j}\gamma_{k})_{ba}
\label{dlxxxi} \ .
\ee
Let $J_s$ be the representation of choice for  the projective module of 
spinors, and $J_T$ for rank 1
tensors. The transform of (\ref{dlxxxi}) onto spinorial modules is:
\be
-\frac{1}{8}c_{ijk}\left[ D^{J_s}(\gamma_j\gamma_k)(D^{J_s})^\dagger
\right]_{\sigma'\sigma}(D^{J_T})^\dagger_{i\rho}\ ,
\label{dlxxxii}\ee
while the torsion--free covariant derivative $\overline\nabla_\rho$ 
acts on a spinor $\alpha^{J_s}$ represented as an element of a 
projective module as follows:
\be
\overline\nabla_\rho\alpha^{J_s}_\sigma=\nabla_\rho\alpha^{J_s}_\sigma-
\frac{1}{8}\alpha^{J_s}_{\sigma'}c_{ij'k'}(D^{J_s}(\gamma_{j'}\gamma_{k'})
(D^{J_s})^\dagger)_{\sigma'\sigma}(D^{J_T})^\dagger_{i\rho} \ .
\label{dlxxxiii}
\ee

\subsection{The Projective Dirac Operator for Spheres}

The equations (\ref{dli}) tell us the invertible transformation of a spinor
field
of \S\;3.3 to an element of a projective module. So we can transform the
Dirac operator $D$ to one acting on this $\cal A$--module. The result is not
illuminating except in special cases like spheres and ${\mathbb C}P^N$, 
so we take them up first.

\noindent a) {\it Even Spheres}

For $G/H=S^{2n}$, we can choose $G=Spin^\cliff(2n+1)=\{g\}$, 
$H=Spin^\cliff(2n)=\{h\}$,
identifying them with the representations given by $\gamma$--matrices,
$Spin^\cliff(2n+1)$ and $Spin^\cliff(2n)$.
We denote the $\gamma$-matrices of $H$ by $\gamma_i,\ i=1,..,2n$, and by
$\gamma=(-i)^n\gamma_1
\ldots\gamma_{2n}$ the additional gamma matrix of $G$, and call them
collectively as $\Gamma_\lambda=(\gamma_i,\gamma),\ \lambda=1,..,2n+1$.
The generators of $H$
are $\Sigma_{ij}=\frac{1}{4i}[\gamma_i,\gamma_j]$, which together
with $\Sigma_{2n+1,i}=\frac{1}{2i}\gamma\gamma_i$ make up the full set
of generators $\Sigma_{\mu\nu}$ of  $G$.

The $\Gamma_\lambda$ transform as vectors under conjugation by $G$. That
lets us introduce coordinate functions $x= (x_\lambda)$ for $S^{2n}$,
starting from an `origine' $x^0=(0,...,0,1)$, as
follows:
\be
\Gamma_\lambda x_\lambda =g \Gamma_{2n+1}g^{-1}\ ,\quad 
g\in Spin(2n+1)\ ,\quad
     x_\lambda x_\lambda=1\ .
\label{dlii}
\ee

We let subscript $A=(\mu\nu),\,\mu>\nu$ stand for either of the
multi-indices $(ij),\ (\alpha\ $ of Sec.2), or
$(2n+1,i),\ (i\ $ of Sec.2).
For $A=(2n+1,i)$, $X_A$ gives back $X_{2n+1,i}\equiv X_i$ of Sec.2, 
which is now torsionless, $G/H$ being symmetric.

Since $\Gamma_{2n+1}$ commutes with $\Sigma_{ij}$, $D_W$ can be written as:
\be
D_W=-i\gamma_i^{\cal R} X_i=[\Gamma_{2n+1},\Sigma_A]^{\cal R} X_A=
[\Gamma_\lambda x_\lambda,\Sigma_A]^{\cal R}X_A\ ,\ \hbox{at}\
x=x^0\ ,
\label{dliv}\ee
while
\be
D=i\Gamma_{2n+1}^{\cal R}D_W=i\Gamma_\lambda^{\cal R} x_\lambda D_W\ ,
\ \hbox{at}\ x=x^0 \ .
\label{dlivb}
\ee

We choose $J$ to correspond to the preceding Clifford representation 
to fix the spinorial projective module. We now show that on this module 
the above Dirac operators have the beautiful forms
\be
{\cal D}_W=i[\Gamma_\lambda^{\cal R} x_\lambda,\Sigma_A^{\cal R}]{\cal J}_A
\quad,\quad
{\cal D}=i\Gamma_\lambda^{\cal R} x_\lambda{\cal D}_W\ ,
\label{dlv}
\ee
 ${\cal J}_A$ being again the total 'orbital' plus 'spin' generators
${\cal L}_A$ and $-\Sigma_A^{\cal R}$ of $G$. The matrices 
$\Gamma_\lambda^{\cal R},\
\Sigma_A^{\cal R}$ act on the index $a$ of the spinor
\be
\chi_a=\sum\xi^K_MD^K_{Mb}(D^\cliff)^\dagger_{ba}
\label{dlvi}
\ee
on the right as in $(\Gamma^{\cal R}_\lambda\chi)_a=
\chi_b(\Gamma_\lambda)_{ba}$.

In fact, if we apply (\ref{dlv}),
since by (\ref{dlviii}) ${\cal J}_A(D^\cliff)^\dagger=0$, we can see that
\be
({\cal D}_W\chi)_a=
i\sum\Big({\cal L}_A(\xi^K_M D^K_{Mb})\Big)(D^\cliff)^\dagger_{bc}
[\Sigma_A,\Gamma_\lambda x_\lambda]_{ca}\ .
\label{dlx}
\ee
Inserting
\beqa
[\Sigma_A,\Gamma_\lambda x_\lambda]&=&
(Ad g)_{AB}D^\cliff(g)[\Sigma_B,\Gamma_{2n+1}]
(D^\cliff(g))^\dagger=\nn\\
&=&i(Ad\,g)_{A,(2n+1,i)}D^\cliff(g)\gamma_i(D^\cliff(g))^\dagger\ ,
\label{dlxi}
\eeqa
we get
\be
({\cal D}_W\chi)_a=-\sum ({\cal L}_A\xi^K_MD^K_{Mb})(Ad\,g)_{A,(2n+1,i)})
(\gamma_i)_{ba'}(D^\cliff(g))^\dagger_{a'a}\ .
\label{dlxii}
\ee
But the right-invariant vector fields  are related to
the left-invariant ones  by eq.(\ref{dliii}), so
\be
({\cal D}_W\chi)_a=-i(X_i\sum\xi^K_MD^K_{Mb})(\gamma_i)_{bc}
(D^\cliff(g))^\dagger_{ca}\ .
\label{dlxiii}
\ee
Writing $\psi_a=\chi_{a'}D^\cliff_{a'a}$, (\ref{dlxiii})
 shows that under ${\cal D}_W\ :
\ \psi_{a}\rightarrow (D_W\psi)_{a}$, which is the action (\ref{dliv}). So
${\cal D}_W$ is equivalent to $D_W$.
In a similar manner ${\cal D}$ is seen to be equivalent to $D$.

When acting on functions on $S^{2n}$, we can use our coordinates to express
the right-invariant vector fields in the form
\be
{\cal L}_{\mu\nu}=-i\Big(x_\mu\frac{\partial}{\partial x_\nu}
-x_\nu\frac{\partial}{\partial x_\mu}\Big)\ ,
\label{dtli}
\ee
and therefore the Dirac operators as
\be
{\cal D}_W=-x_\mu\Gamma^{\cal R}_\nu({\cal L}_{\mu\nu}-\Sigma^{\cal
R}_{\mu\nu})\quad ,
\quad {\cal D}=-\Sigma^{\cal R}_{\mu\nu}{\cal L}_{\mu\nu}+n  \ .
\label{dtlii}
\ee

To determine the spectrum and eigenspinors of the Dirac operator
we need to be more explicit about the group $Spin(2n+1)$. It has rank
$n$, and IRR's that can be labeled by the components of the highest weight
$(m_1,...,m_n)$, with the $m_i$'s all integers or all half integers,
and $m_1\ge m_2\ge...\ge m_n\ge 0$. The Clifford representation $Spin^\cliff$
 has highest weight $(\frac{1}{2},.., \frac{1}{2})$,
dimension $2^n$ and quadratic Casimir operator $C_2(\cliff)\equiv 
C_2(Spin^\cliff)=\frac{1}{2}\Sigma_{\mu\nu}\Sigma_{\mu\nu}=
\frac{1}{4}n(2n+1)$. We indicate by $L$ an IRR associated with the
set $I_0$ of \S 2.2; it has highest weight $(l,0,..,0)$, 
where $l$ is an integer, and dimension and quadratic Casimir operator
\be
d(L)= \frac{2l+2n-1}{l+2n-1}\frac{(l+2n-1)!}{l!(2n-1)!}
\quad,\quad C_2(L)=l(l+2n-1)  \ .
\label{dtliv}
\ee
The final piece of required information is
\be
L\otimes \cliff =(l+\frac{1}{2},\frac{1}{2},..,\frac{1}{2})\oplus
(l-\frac{1}{2},\frac{1}{2},..,\frac{1}{2})
\label{dtlv}
\ee
with
\be
d(j,\frac{1}{2},..,\frac{1}{2})=
2^n\frac{(j+2n-\frac{3}{2})!}{(j-\frac{1}{2})!(2n-1)!}\ ,\quad
C_2(j,\frac{1}{2},..,\frac{1}{2})=
j(j+2n-1)+\frac{1}{2}(n-1)(n-\frac{1}{2})\ .
\label{dtlvi}
\ee
With this background it is easy to show that the
eigenspinors of $\cal D$ are of the form
\be
\chi^{JL}_a=\sum \xi^{JL}_M <JM|LN,\cliff\; a>D^L_{Ni_0}
\ \ \hbox{with}\ J=(l\pm\frac{1}{2},\frac{1}{2},..,\frac{1}{2})   \ ,
\label{dtlvii}
\ee
where $M,N$ and $i_0$ , and $a$ label vectors in the IRR's $J,L$ and 
$\cliff$.
 In fact
\be
({\cal D}\chi)_a=\sum\xi^{JL}_M <JM|LN',\cliff\; a'>
(\Sigma_{\mu\nu})_{a'a}(\Sigma_{\mu\nu}^L)_{N'N}D^L_{Ni_0}+n\chi_a\ ,
\label{dtlviii}
\ee
where $\Sigma_{\mu\nu}^L$ is the representative of ${\cal L}_{\mu\nu}$ in the
IRR $L$.`Completing the square' in this equation, one finds
\be
 <JM|LN',\cliff\; a'>
 (\Sigma_{\mu\nu})_{a'a}(\Sigma_{\mu\nu}^L)_{N'N}=
 (C_2(J)-C_2(L)-C_2(\cliff)<JM|LN,\cliff\; a> \ .
 \label{dtlix}
\ee
Using the expressions for the various quadratic Casimir operators,  the
eigenvalues corresponding to the the eigenspinors (\ref{dtlvii}) are
found to be
\be
 \rho=\pm(j+n-\frac{1}{2})  \quad ,\ \hbox{for}\ j=l\pm\frac{1}{2}\ .
 \label{dtlx}
\ee
\bigskip

\noindent b) {\it Odd Spheres}

An odd sphere $S^{2n-1}=SO(2n)/SO(2n-1)$ differs from an even sphere
$S^{2n}$ in important details. The Clifford algebra $\cliff(2n-1)$
has two inequivalent $2^{n-1}$-dimensional representations, with
$(-i)^{n-1}\gamma_1...\gamma_{2n-1}=\I$ and 
$(-i)^{n-1}\tilde\gamma_1...\tilde\gamma_{2n-1}=-\I$; we may take 
$\tilde\gamma_i=-\gamma_i$, which makes clear that they give a single
IRR's of $Spin(2n-1)$, with generators $\frac{1}{4i}[\gamma_i,\gamma_j]$.
They do give however two inequivalent IRR's of $Spin(2n)$, with generators
$(\frac{1}{4i}[\gamma_i,\gamma_j],-\frac{1}{2}\gamma_i)$ and 
$(\frac{1}{4i}[\gamma_i,\gamma_j],\frac{1}{2}\gamma_i)$, let us label them 
$\cliff^+$ and  $\cliff^-$.

For covariance it is better to put  these two representations together
and work with  the $2^n$-dimensional $\Gamma_\mu,\ \mu=1,...,2n$, 
built from the $\gamma_i$-s as indicated in (\ref{dxxiva}); that
particular construction gives 
\be
\Gamma_{2n+1}=(-i)^n\Gamma_1...\Gamma_{2n}=
\left(\begin{matrix}\I &0\\0&-\I \end{matrix}\right)\ ,\quad 
\tilde\Gamma_{2n-1}\equiv(-i)^{n-1}\Gamma_1...\Gamma_{2n-1}=
\left(\begin{matrix}0&\I \\ \I &0\end{matrix}\right)\ .
\nn\ee
For $\cliff(2n-1)$,
 $\Gamma_\mu$  splits into the two inequivalent IRR's
$\frac{1}{2}(\I \pm\tilde\Gamma_{2n-1})\Gamma_j\equiv
\Gamma_j^{(1,2)}$, $1\le j\le 2n-1$.
The corresponding generators of $Spin^\cliff(2n)$ are:
\be
\Sigma_{\mu\nu}=\frac{1}{4i}[\Gamma_\mu,\Gamma_\nu]=\Big\{
\Sigma_{ij}=\left(\begin{matrix}\frac{1}{4i}[\gamma_i,\gamma_j]&0\\ 
0&\frac{1}{4i}[\gamma_i,\gamma_j]\end{matrix}\right)\ ;\ 
\Sigma_{2n,i}=\left(\begin{matrix}-\frac{1}{2}\gamma_i&0\\
0&\frac{1}{2}\gamma_i\end{matrix}\right)\Big\}\ ,
\label{dlxiv}
\ee
and give, as expected, the direct sum $\cliff^+\oplus\cliff^-$; 
for the group elements of $Spin(2n)$
we have $g=\left(\begin{matrix}D^{\cliff^+}&0\\ 
0&D^{\cliff^-}\end{matrix}\right)$, split by the projectors
$\frac{1}{2}(\I \pm\Gamma_{2n+1})$.

  Spinors  carry the direct sum of these two IRR's on their index, and
we can use either of the Dirac operators
\be
 D^{(1,2)}_W= 
 \frac{1}{2}(\I \pm\tilde\Gamma_{2n-1} )^{\cal R} (-i 
\,\Gamma_i^{\cal R}) X_i\ , 
\label{dlxiva}
\ee
 or else we can accept fermion doubling and work with
$D^{(1)}_W + D^{(2)}_W$. 

There is no chirality in odd dimensions, but $\Gamma_{2n}$ plays a
role in space(time)-reflection, and can be used to give  Dirac operators 
equivalent to $D^{(1,2)}_W$ \cite{peter}:
\be
D^{(1,2)}=e^{i\Gamma_{2n}^{\cal R}\pi/4}D^{(1,2)}_W
e^{-i\Gamma_{2n}^{\cal R}\pi/4}=\frac{1}{2}(\I \mp\Gamma_{2n+1}^{\cal
R})\,\Gamma_{2n}^{\cal R}\,\Gamma^{\cal R}_i\,X_i\ .
\label{dlxivb}
\ee
We can introduce coordinates for $S^{2n-1}$, starting from $x^0=(0,..,1)$, 
by
\be
\Gamma_\lambda x_\lambda= g\Gamma_\lambda x_\lambda^0 g^{-1}=
g\Gamma_{2n}g^{-1} \ ,\quad x_\lambda x_\lambda=1\ .
 \label{dlxv}
\ee
 Hence at $x=x^0$,
\beqa
 D^{(1,2)}_W&=&p(x^0)_{(1,2)}^{R}[\Gamma_\lambda
x_\lambda^0,\Sigma_A]^{\cal R}X_A\ , \quad 
D^{(1,2)}=\frac{1}{2}(\I \mp\Gamma_{2n+1}^{\cal R})i
\Gamma_\lambda^{\cal R} x_\lambda^0
[\Gamma_\lambda x_\lambda^0,\Sigma_A]^{\cal R}X_A ,
 \nonumber\\
 p(x^0)_{(1,2)}&=&\frac{1}{2}(\I \pm\tilde\Gamma_{2n-1})=
 \frac{1}{2}\left(\I \pm\frac{(-i)^{n-1}}{(2n-1)!}
\epsilon_{\mu_1\ldots\mu_{2n}}
 \Gamma_{\mu_1}\ldots\Gamma_{\mu_{2n-1}}x^0_{\mu_{2n}}\right)\ .
 \label{dlxvi}
\eeqa
Their covariant forms follow:
\beqa
{\cal D}^{(1,2)}_W&=&
p(x)^{\cal R}_{(1,2)}i\,[\Gamma^{\cal R}_\lambda x_\lambda,
\Sigma_A^{\cal R}] {\cal J}_A\ , \quad {\cal D}^{(1,2)}=
-\frac{1}{2}(\I \mp\Gamma_{2n+1}^{\cal R})\Gamma_\rho^{\cal R} x_\rho
[\Gamma_\lambda^{\cal R} x_\lambda,\Sigma_A^{\cal R}]{\cal J}_A  \ ,
\nonumber\\ p(x)_{(1,2)}&=&\frac{1}{2}\left( \I \pm\frac{(-i)^{n-1}}
{(2n-1)!}\epsilon_{\mu_1...\mu_{2n}}
\Gamma_{\mu_1}...\Gamma_{\mu_{2n-1}}x_{\mu_{2n}}\right)=g\,p(x^0)\,g^{-1}\ .
\label{dlxvii}
\eeqa
${\cal J}_A$ is defined as before.

Proceeding as we did for even spheres, with ${\cal L}_{\mu\nu}$ as in 
eq.(\ref{dtli}) the Dirac operators can be rewritten in the form
\be
{\cal D}^{(1,2)}_W=-p(x)^{\cal R}_{(1,2)}
x_\mu\Gamma^{\cal R}_\nu({\cal L}_{\mu\nu}-\Sigma^{\cal R}_{\mu\nu})
\quad,\quad  {\cal D}^{(1,2)}=\frac{1}{2}(\I \mp\Gamma_{2n+1}^{\cal R})
(-\Sigma^{\cal R}_{\mu\nu}{\cal L}_{\mu\nu}+n-\frac{1}{2})\ .
\label{dlxviii}
\ee
Given their form,
it is easy to find one set of eigenvalues and eigenspinors for the Dirac
operators ${\cal D}^{(1,2)}$, by the same argument that led us to 
eq.(\ref{dtlvii}). The IRR's of
$Spin(2n)$ are labeled by highest weights
$(m_1,..,m_n),\ m_1\ge m_2...\ge |m_n|\ge 0$ with the $m_i$ all integers or
all half integers. The two $2^{n-1}$-d spinor representations $\cliff^\pm$
have $(\frac{1}{2},..,\frac{1}{2},\pm\frac{1}{2})$, with quadratic Casimir 
$C_2(\cliff^\pm)=\frac{1}{2}\Sigma_{\mu\nu}\Sigma_{\mu\nu}=
\frac{1}{4}n(2n-1)$. The IRR's associated with the set $I_0$ of section 2.2
are $L=(l,0,..)$, with dimension 
$d(L)=\frac{(l+n-1)\,(l+2n-3)!}{(n-1)\,(2n-3)!\,l!}$,
and quadratic Casimir $C_2(L)=l(l+2n-2)$. Finally
\be
 L\otimes  \cliff^\pm=
 (l+\frac{1}{2},\frac{1}{2},..,\pm \frac{1}{2})\oplus(l-\frac{1}{2},
\frac{1}{2},..,\pm\frac{1}{2})\ .
 \label{dlxviiia}
\ee
For these last representations we have $C_2(j,\frac{1}{2},\dots ,
\pm\frac{1}{2})=(j-\frac{1}{2})(j+2n-\frac{3}{2})+\frac{1}{4}n(2n-1)$.

With all this information, the analogues of (\ref{dtlviii}),(\ref{dtlix}) 
give for the eigenvalues
\be
\rho_\pm=\pm(j_\pm+n-1)\quad,\hbox{ with}\ j_\pm=l\pm\frac{1}{2}\ .
\label{dlxviiib}
\ee
There are eigenstates for each  of the two inequivalent 
representations of the Clifford algebra. They are given by
\beqa
\chi_{1,a\pm}^{JL}&=&\big(\ 0\quad ,\ \sum\xi^{JL}_M<
(j_\pm,\frac{1}{2},..,-\frac{1}{2}),M|
L,N';\cliff^-\;a\;>D^L_{N'i_0}\big)\ ,\nn\\
\chi_{2,a\pm}^{JL}&=&\big(\sum\xi^{JL}_M
<(j_\pm,\frac{1}{2},..,\frac{1}{2}),M|
L,N';\cliff^+\;a\;>D^L_{N'i_0}\ ,\quad 0\ \big)
\ .\label{dtlviiib}
\eeqa
However, from eq.(\ref{dlxivb}), we have
\be
\Gamma_{\lambda}^{\cal R}x_\lambda\, {\cal D}^{(1,2)}=-{\cal D}^{(2,1)}
\,\Gamma_{\lambda}^{\cal R}x_\lambda\ ,
\label{dtlviiic}
\ee
and this implies that there is another set of eigenvectors, with the 
same  eigenvalues, given by
\beqa
\tilde\chi_{1,a\pm}^{JL}=\Gamma_\lambda x_\lambda\,\chi^{J,L}_{2,a\mp}
&=&\big(\ 0\quad ,\ \sum\xi^{JL}_M<(j_\mp,\frac{1}{2},..,\frac{1}{2}),M|D^J|
L,i_0;\cliff^+\;\alpha\;>D^{\cliff^-\dagger}_{\alpha\, a}\big)\ ,\nn\\
\tilde\chi_{2,a\pm}^{JL}=\Gamma_\lambda x_\lambda\,\chi^{J,L}_{1,a\mp}
&=&\big(\sum\xi^{JL}_M<(j_\mp,\frac{1}{2},..,-\frac{1}{2}),M|D^J|
L,i_0;\cliff^-\;\alpha\;>D^{\cliff^+\dagger}_{\alpha\, a}\ ,\ 0\ \big)
\ .\nn\\ \label{dtlviiid}
\eeqa

\subsection{The Projective Dirac Operators on ${\mathbb C}P^N$}

For reasons of brevity, we focus on ${\mathbb C}P^2$, a case we have already
treated in \cite{cp2}.  ${\mathbb C}P^2$ is $SU(3)/U(2)$. If $\lambda_\alpha$
are the Gell-Mann matrices, it is the orbit of $\lambda_8$ under $SU(3)$:
\be
 {\mathbb C}P^2\ :\quad \{g\lambda_8 g^{-1}\ ,\ g\in SU(3)\}\ .
\label{dlxix}
\ee
Writing $g\lambda_8 g^{-1}=\lambda_A\xi_A$ analogously to
(\ref{dlii}), we
can regard those $\xi\in{\mathbb R}^8$ given by(\ref{dlxix}),
as points of $ {\mathbb C}P^2$.
The stability group at $\lambda_8$, or equally well at $\xi^0=(0,\ldots,0,1)$
is $U(2)$. Its generators are $\lambda_1,\lambda_2,\lambda_3,\lambda_8$.

If we can achieve a covariant--looking form for $D$ and $D_W$ looking like
(\ref{dliv}), (\ref{dlivb}),
we can find covariant $\cal D$ and ${\cal D}_W$. Towards this end we introduce
the Clifford algebra with eight generators $\gamma_A$. They can be transformed
by the adjoint representation of $SU(3)$ without disturbing their
anticommutators:
\be
\gamma'_A=Ad\;g_{AB}\gamma_B\ \rightarrow\ \{\gamma'_A,\gamma'_B\}=
2\delta_{AB}\ .
\label{dlxx}
\ee
The generators $\Sigma_A$ in this representation can actually be written
using $\gamma_A$:
\be
\Sigma_A=\frac{1}{4i}f_{ABC}\gamma_B\gamma_C\ .
\label{dlxxi}
\ee

Consider the action $\gamma_A\rightarrow [\Sigma_8,\gamma_A]$ of $\Sigma_8$
on $\gamma_A$. For this action,
the eigenvalues of $\Sigma_8$ are $\pm\frac{\sqrt 3}{ 2}$ and $0$.
The $0$ eigenvalues are for $\gamma_A$ with $ A=1,2,3,8$, thus:
\beqa
[\Sigma_8,[\Sigma_8,\gamma_A]]&=&0\quad \hbox{if}\quad A=1,2,3,8\ ,
\nonumber\\    &=&\frac{3}{4}\gamma_A\quad \hbox{if}\quad A=4,5,6,7\ .
\label{dlxxii}
\eeqa
This lets us write the Dirac operator in 'covariant' form
\be
D_W=-i\,\frac{4}{3}\,
[\Sigma\cdot\xi^0,[\Sigma\cdot\xi^0,\gamma_A]]^{\cal R} X_A\quad,\quad
(X_Af)(g)=\frac{d}{dt}f(ge^{it\Sigma_A})|_{t=0}  \ .
\label{dlxxiii}
\ee
The role of $\Sigma_A$ and $\gamma_A$ are reversed here for covariantization
as compared to spheres.

For the projective module, for the representation $D^J$, we have the
one given by $\Sigma_A$. It is $2^{8/2}=16$- dimensional. The transform
${\cal D}_W$ of $D_W$ onto this module is immediate:
\be
{\cal D}_W=- 
\frac{4}{3}\,[\Sigma\cdot\xi,[\Sigma\cdot\xi,\gamma_A ]]^{\cal R}{\cal J}_A
\ .\label{dlxxiv}
\ee
In addition to $D_W$ we can also write the Dirac operators
\be
D'=-i\,\frac{2}{\sqrt 3}\,[\Sigma\cdot\xi^0,\gamma_A]X_A\quad,\quad D=
i\Gamma(\xi^0)D_W
\quad,\ \Gamma(\xi^0)=-\gamma_4\gamma_5\gamma_6\gamma_7\ .
\label{dlxxv}
\ee
$D'$ becomes ${\cal D}'=-i\,\frac{2}{\sqrt 3}[\Sigma\cdot\xi,\gamma_A]J_A$
on the projective module.
To find ${\cal D}$ we need to find the chirality operator
$\Gamma(\xi)$ for all $\xi$. This is in \cite{cp2} and is just
\be
\Gamma(\xi)=-\frac{1}{4!}\epsilon_{ABCD}(\xi)\gamma_A\gamma_B\gamma_C\gamma_D
\quad,\quad
\epsilon_{ABCD}(\xi)=4(ad\;\Sigma\cdot\xi)_{[AB}(ad\;\Sigma\cdot\xi)_{CD]}
\label{dlxxvi}\ee
($ [\ ]=$antisymmetrization).
We cannot have an $a$ in $\chi_a$ take values from 1 to 16: that would give
4 spinors. We must have it taking just 4 values and carrying the representation
of just $\gamma_A\ ,\ A=4,5,6,7$. The explanation of how this is done
takes up some space in \cite{cp2}.

\section{On Riemannian Structure and Gravity.}
\setcounter{equation}{0}

An inverse metric $(\eta^{ij})$ is a symmetric nondegenerate field, which
defines a map ${\cal T}^{(1)}\otimes{\cal T}^{(1)}\rightarrow{\cal T}^{(0)}$
via $f\otimes f'\rightarrow \eta^{ij}f_if'_j$. As the $f$'s transform
by $Ad_{G/H}$
under $g\rightarrow gh$, $(\eta^{ij})$ transform by the product
$Ad^{-1}_{G/H}\otimes Ad^{-1}_{G/H}$ of its contragradient representation. 
Or, the metric $(\eta_{ij})$ itself transforms by 
$Ad_{G/H}\otimes Ad_{G/H}$.

A particular metric is $(\hat\delta_{ij})$, where $\hat\delta_{ij}(g)$ is
$\delta_{ij}$ ($\delta$= Kronecker $\delta$). The torsion--free covariant
derivative compatible with $\hat\delta$ is $\overline X$:
\be
\overline X\hat\delta=0\ .
\label{dlxxxiv}
\ee
The corresponding curvature tensor of $G/H$ can be calculated in terms of the
structure constants of $G$. From their form we have that for any vector field
$f_i$ tangent to $G/H$,
\be
[\overline X_i,\overline X_j]=R_{ijkl}\,f_l=\big( c_{ij\alpha}c_{\alpha kl}+
\frac{1}{4}(2\,c_{ijk'}c_{k'kl}-c_{ikk'}c_{k'jl}-c_{kjk'}c_{k'il})\big)f_l\ .
\label{dlxxxiva}
\ee
The scalar curvature is then $R=R_{ijij}=c_{ij\alpha}c_{\alpha ij}+
\frac{1}{4}c_{ijk'}c_{k'ij}$. For $S^n=Spin(n+1)/Spin(n)$, we found in
section 5.3 that in a Clifford representation, $[\Sigma_{n+1,i},
\Sigma_{n+1,j}]=i\Sigma_{ij}$. So, with the correspondences 
$i\leftrightarrow (n+1,i),\ \alpha\leftrightarrow (i,j)$ we have that
$c_{ijk}=0$, and the curvature is $R=n(n-1)$. 

A more general $H$-invariant metric $\eta$ can be defined as
follows. Let us decompose $\underline{G/H}$ into irreducible subspaces
under $Ad_{G/H}$ and let $\{S^{(\sigma)}_m\}$ be a basis for the unitary
irreducible representation $\sigma$ such that
\be
\tr S^{(\sigma)^\dagger}_mS^{(\sigma')}_n = c
\delta_{\sigma\sigma'}\delta_{mn}\ .
\label{dlxxxivb}
\ee
(Here $\sigma$ and $\sigma'$ can be equivalent representations). Let 
$X^{(\sigma)}_m$ be the corresponding (in general complex) vector
field, it is a linear combination of $X_i$. Then a general
$H$-invariant metric $\eta$ on vector fields $X^{(\sigma)\dagger}_m$
and $X^{(\sigma')}_n$ is the constant function defined by
\be
\eta(X^{(\sigma)\dagger}_m,X^{(\sigma')}_n)(g) = \lambda_\sigma
\delta_{\sigma\sigma'}\delta_{mn}\ ,\ \lambda_\sigma\ {\rm a\
positive\ constant}\ ,
\label{dlxxxivc}
\ee
independent of $g$. Such metrics are essential for certain K\" ahler
structures as we shall see in Section 7.

The general covariant differential $\nabla$ can be defined in the usual way:
\be
\nabla_i\eta_{jk}=\overline X_i\eta_{jk}+\Gamma^{j'}_{ij}\eta_{j'k}+
\Gamma^{k'}_{ik}\eta_{jk'}\ .
\label{dlxxxv}
\ee
The formula shows that $\Gamma$ transforms by $Ad_{G/H}\otimes Ad_{G/H}
\otimes Ad^{-1}_{G/H}$ under the structure group $H$. As $\overline X$ is
torsion--free, so is $\nabla$ if as usual $\Gamma^{k}_{ij}=\Gamma^{k}_{ji}$.
A standard calculation gives the metric--compatible  torsion--free
$\nabla$, its $\Gamma$ being given by
\be
\Gamma^k_{ij}=-\frac{1}{2}\eta^{kk'}\left(\overline X_i\eta_{jk'} +
 \overline X_j\eta_{ik'}-\overline X_{k'}\eta_{ij}\right)\ .
\label{dlxxxvi}
\ee
These $\Gamma^k_{ij}$ are not Christoffel symbols, for example they vanish
if $c_{ijk}=0$. Christoffel symbols are defined with respect to some local 
coordinates $x^a$ on $G/H$.

Next, introduce $|G/H|$-beins or soldering forms $e^a_i$ such that
\be
\eta_{ij}=\eta(X_i,X_j)=e^a_ie^b_j\eta_{ab}\ .
\label{dlxxxvii}
\ee
The Christoffel symbols are defined from $\eta_{ab}$ in the usual way.

The spin connection is defined by:
\be 
\nabla_ie^a_j=\overline X_i e^a_j+\Gamma^{k}_{ij}e^a_k+
e^b_j(\omega_i)_{ba}=0\ ,
\label{dlxxxviii}
\ee
where $(\omega_i)_{ba}=-(\omega_i)_{ab}$ and transforms as a tensor field
in $i$ under $H$. The solution for $\omega_i$ is standard:
\be
(\omega_i)_{ca}=-E^j_c[\overline X_ie^a_j+\Gamma^k_{ij}e^a_k]\ ,\quad
E_c^j e^a_j=\delta^a_c \quad\hbox{or}\quad E^j_c=\eta^{jk}e^a_k\eta_{ac}\ .
\label{dlxxxix}
\ee

The covariant derivative on spinors $\psi$ is given by
\be
(\nabla_i\psi)_a=(\overline X_i\psi)_a-\frac{1}{4}(\omega_i)_{cd}
(\gamma_c\gamma_d\psi)_a\ .
\label{dxca}
\ee

The Dirac operator in the presence of a gravity field $(\eta_{ij})$ is thus:
\be
D=\eta^{ij}e_j^a\gamma_a \nabla_j\ .
\label{dxc}
\ee

All this stuff is very natural. It remains to transport it to projective
modules. In the module picture $\eta_{ij}$ gets transformed to
\be
G_{\lambda\rho}=\eta_{ij}(D^J)^\dagger_{i\lambda}(D^J)^\dagger_{j\rho}\ ,
\label{dxci}
\ee
while $\eta^{ij}$ becomes
\be
G^{\lambda\rho}=\eta^{ij}D^J_{\lambda i}D^J_{\rho j}\ .
\label{dxcib}
\ee
The projector for the module is
\be
P^\lambda_{\ \sigma}=G^{\lambda\rho}G_{\rho\sigma}=D^J_{\lambda i}
(D^J)^\dagger_{i\sigma}\ .
\label{dxcii}
\ee

The projective module analogue of $\overline X_i$ is
the $\overline \nabla_\rho$ defined in section 5.2.  Adding the action of
\be
\Gamma^\nu_{\lambda\mu}=
\Gamma^k_{ij}(D^J)^\dagger_{i\lambda} (D^J)^\dagger_{j\nu}
(D^J)^\dagger_{k\nu}
\label{dxciv}
\ee
to $\overline\nabla_\rho$
defines the action of $\nabla_\rho$, the metric-compatible torsion-free
covariant derivative on tensors ($\nabla_\rho G_{\mu\nu}=0$).

The action of $\nabla_\rho$ on spinorial modules follows from (\ref{dxca}). We
let
${\cal J}_\rho$ be the total angular momentum for the representation $J_S$
chosen for spinors, and
\beqa
{\cal C}^{(S)}_{\rho\lambda\sigma}&=&D^{J_S}_{\lambda b}
(c_{ijk}\gamma_j\gamma_k)_{ba}(D^{J_S})^\dagger_{a\sigma}
(D^{J_T})^\dagger_{i\rho}  \ ,   \nonumber\\
\Omega_{\rho\lambda\sigma}&=&D^{J_S}_{\lambda b}
((\omega_i)_{jk}\gamma_j\gamma_k)_{ba}(D^{J_S})^\dagger_{a\sigma}
(D^{J_T})^\dagger_{i\rho}  \ ,     \nonumber\\
\chi_\sigma&=& \psi_a(D^{J_S})^\dagger_{a\sigma}\ .
\label{dxcv}
\eeqa
Then, as can easily be shown from (\ref{dlxxxiii}) above,
\be
(\nabla_\rho\chi)_\sigma=-i({\cal J}_\rho\chi)_\sigma-
\chi_\lambda\left(\frac{1}{8}{\cal C}^{(S)}+\frac{1}{4}
\Omega\right)_{\rho\lambda\sigma}\ .
\label{dxcvi}
\ee

\section{Complex Structures and K\"ahler Manifolds}
\setcounter{equation}{0}

In favourable circumstances, we can push this program ahead and define
more refined ideas like complex and K\"ahler structures on tensors
${\cal T}^{(n)}$ and
on their projective modules. We indicate how to treat them briefly.

We consider adjoint orbits only for $G/H$. Thus let
$\underline{k}$ be a fixed element of $\underline G$ from the
Cartan subalgebra $C(\underline G)$, and $H$ its stability group:
\be
H\ =\ <h\in G\ :\ h\underline k h^{-1}=\underline k>\quad,\quad
[\underline{k}, T(\alpha)]=0\ \forall\ \alpha\quad,\quad
[\underline{k}, S(i)]\ne 0\ \forall\ i\ .
\label{dxcvii}
\ee
The Cartan subalgebra of $\underline H$,
 $C(\underline H)=C(\underline G)$, since any
element of $\underline G$ which commutes with $\underline k$ is in
$\underline H$.
The manifold $G/H$, being an adjoint orbit of the simple Lie
group  $G$, has even dimension. These observations have the following 
implications. 

Consider the eigenvalue equation 
\be [\underline k,E_a]=\lambda_aE_a\ .
\label{dxcviia} 
\ee 
Then $\lambda_a\ne 0$. The $E_a$ will be of
the form $\sum_i\xi_{ai}S(i),\ \xi_{ai}\in{\mathbb C}$, and span
the complexification $(\underline{G/H})_c$ of $(\underline{G/H})$.

By (\ref{dviia}), $(\underline{G/H})_c$ is invariant under the adjoint
action of $\underline k$. Also, as $Ad_{\underline{G/H}}$ is a
real, orthogonal representation, the eigenvalues $\lambda_a$ are
real, while of course the $S(i)$ are hermitean. 
So the adjoint of (\ref{dxcviia})
shows that $E^\dagger_a$ corresponds to the eigenvalue
$-\lambda_a$, and that each positive eigenvalue is paired with a
negative one. The eigenvalues $\lambda_a$ may be degenerate. 

We choose  $E_a,\ a=1,...,\half(|G|-|H|)$,
to be solutions of (\ref{dxcviia}) with $\lambda_a>0$,
$E_{-a}=E^\dagger_a$, and the normalization 
\be 
\tr\,
E_a\,E_b=\delta_{a+b,0}\ . 
\label{dxcviib} 
\ee 
So, if $E_a=\xi_{a\,i}S(i)$ (for both signs of $a$), 
$\xi_{-a\,i}=\xi^*_{a\,i}$. We choose $c=1$ in (\ref{di}) and
(\ref{diii}). Then  (\ref{diii}) and  (\ref{dxcviib}) show that the 
matrix $\{\xi_{a\,i}\}$ is unitary as well.

Let $(\underline{G/ H})_c^\pm$ denote the span of the
eigenvectors $E_{\pm |a|}$ (where note that $|a|>0$). The subspaces
$(\underline{G/H})_c^\pm$ are of precisely the same dimension
and
\be
(\underline{G/H})_c=(\underline{G/H})_c^+\oplus
(\underline{G/H})_c^-
\label{dxcviic}
\ee
The elements $E^+=\sum_i\xi^iS(i)\in(\underline{G/H})_c^+$
generate vector fields $X^+=\xi^iX_i$ which we define to be 
holomorphic. Let
${\cal H}^+$ denote the space of holomorphic vector fields.
Likewise $(\underline{G/H})_c^-$ gives rise to the space ${\cal H}^-$  of
antiholomorphic vector fields. This splitting of the space of fields
$\cal H$ as the direct sum ${\cal H}^+\oplus{\cal H}^-$ gives us the
complex structure. The $(1,1)$ tensor $J$ of complex analysis is $\pm
i$ on ${\cal H}^\pm$: for a vector field $X=\xi^iX_i,\ JX=\xi^j\sum_{a>0}
i(\xi^*_{aj}\xi_{ai}-\xi_{aj}\xi_{ai}^*)X_i$.

This complex structure is K\"ahler. To show it, let us introduce
the Maurer-Cartan forms $\theta^A$, defined by
$g^{-1}dg=i\Sigma_A\theta^A$, or, setting $c=1$ in (\ref{di}), by 
\be
\theta^A=-i\tr\,\Sigma_A\,g^{-1}\,dg\ . \label{dgxcviid} 
\ee 
They
are dual to the vector fields $X_A=(X_i,\,X_\alpha)$, so that for
example $\theta^i(X_j)=\delta_{ij}$, and fulfill \be
d\theta^A=-\frac{i}{2}\tr(\Sigma_A[\Sigma_B,\Sigma_C])\theta^B\wedge\theta^C=
\half c_{BCA}\theta^B\wedge\theta^C\ . 
\label{dxcviie} 
\ee 

Consider the particular Maurer-Cartan form \be
\Theta=-i\tr\,\underline k\,g^{-1}\,dg=\tr\,(\underline
k\,\Sigma_A)\theta^A=\tr\big(\underline k\,T(\alpha)\big)\theta^\alpha\ .
\label{dxcviif} 
\ee 
In the last step we used (\ref{div}). For $d\Theta$ we have 
\be 
d\Theta=-\frac{i}{2}\tr\big(\underline k\,
[\Sigma_B,\Sigma_C]\big) \theta^B\wedge\theta^C\ . 
\label{dxcviig} 
\ee
But using (\ref{dxcvii}) we have $\tr\underline
k\,[T(\alpha),T(\beta)]=\tr [\underline k,T(\alpha)]T(\beta)=0$ and
$\tr\underline k\,[T(\alpha),S(i)]=\tr[\underline 
k,T(\alpha)]\,S(i)=0$. Therefore 
\be
d\Theta=-\frac{i}{2}\tr\big(\underline
k\,[S(i),S(j)]\big)\theta^i\wedge\theta^j\ .
\label{dxcviih} 
\ee
Remembering that the matrix
$\{\xi_{a\,i}\}$ (with $a$ of both signs) is unitary, we may set
\be
S(i)\,\theta^i=E_a\,\xi^*_{a\,i}\,\theta^i\equiv E_a\,\theta^a
\label{dxcviij}
\ee
 (with implied sum over $a$ of both signs), and rewrite (\ref{dxcviih}) as
\beqa 
d\Theta &=& -\frac{i}{2}(\tr\underline
k[E_a,E_b])\theta^a\wedge\theta^b \nonumber\\
&=& -\frac{i}{2}\lambda_a(\tr\,E_a\,E_b)\theta^a\wedge\theta^b=
-i\sum_{a>0}\lambda_a\theta^a\wedge\theta^{-a}\ , 
\label{dxcviik}
\eeqa 
where we have used (\ref{dxcviia}) and (\ref{dxcviib}). The vector fields
$X_a=\xi_{ai}X_i$ are dual to $\theta^a$:
$\theta^a(X_b)=\delta_{a\,b}$. 
Consequently, the two-form $\Omega=d\Theta$ can be specified by
\be 
\Omega(X_a,X_b)=d\Theta(X_a,X_b)=-i\lambda_a\delta_{a+b,0}\ .
\label{dxcviil}
\ee 
Since all $\lambda_a\neq 0$, it follows from (\ref{dxcviik}) that
$\Omega$ is a symplectic (i.e. closed and non-degenerate) form on
$G/H$. 
It has been extensively discussed in \cite{bal} where its physical 
implications are also explained. It fulfills the K\"ahlerian condition 
\be
\Omega(JX_a,JX_b)=\Omega(X_a,X_b)\ . 
\label{dxcviim} 
\ee
 For vector fields $X=\xi^iX_i,\;Y=\eta^iX_i$, we have $\Omega(X,Y)=
\sum_{a>0}(i\lambda_a)(\xi_{ai}\xi^*_{aj}-\xi_{aj}\xi_{ai}^*)\xi^i\eta^j$.

The K\"ahler metric $\eta$ on vector fields $(X_a,\,X_b)$ is given by 
\be
\eta(X_a,X_b)=\Omega(JX_a,X_b)=|\lambda_a|\delta_{a+b,0}\ . 
\label{dxcviin} 
\ee
The Levi-Civita connection corresponding to this metric is the 
torsion-less connection compatible with $\eta$. Its coefficients are
given by the formula (\ref{dlxxxvi}):
\be \frac{1}{2}c_{abc} +\Gamma^c_{ab} = \frac{1}{2}c_{abc} -
\frac{|\lambda_a|-|\lambda_b|}{2|\lambda_c|}c_{abc}\ .
\label{6.15}
\ee
Note from (\ref{dxcviib}) and (\ref{dvib}) that $c_{abc}=\tr [E_a,E_b]
E_{-c}$. So we have the symmetries 
\be 
c_{abc} = c_{b,-c,-a} = c_{-c,a,-b}\ . 
\label{c.abc} 
\ee
Also from $\tr [\underline{k},[E_a,E_b]E_{-c}]=0$ and (\ref{dxcviia})
we have that 
\be
c_{abc}\ {\rm and}\ (\ref{6.15})\, =\, 0\ {\rm if}\ 
\lambda_a+\lambda_b -\lambda_c\neq 0\ .
\label{c.abc0} 
\ee

Finally, we shall show that the  K\"ahler metric on $G/H$ can be
derived from a  K\"ahler potential $\Phi_\zeta$. It is a function on 
$G/H$ and depends on a parameter $\zeta$. It has the property
\be
X_aX_{-b}\Phi_\zeta = \eta(X_a,X_{-b})
\label{6.16}
\ee
for $\lambda_a$ and  $\lambda_b$ of the same sign and $|\lambda_a| 
\leq |\lambda_b|$. The ordering is needed because of the torsion term
in (\ref{dvib}). It can be discarded when the torsion is zero, that
is for symmetric spaces. Note that $\Phi_\zeta$ can in general be only 
locally defined on $G/H$. 
 
The construction of $\Phi_\zeta$ involves the member of a specific 
class of unitary representations $\Sigma_K:g\to\Sigma_K(g)$ of $G$. Let
$\sigma_K$ be the associated representation of $\underline G$.  Any
such representation contains a normalized highest weight vector $|K>$ with
eigenvalue $K$ for $\sigma_K(\underline{k})$, which is annihilated by 
the orthogonal complement of $\underline k$ in  $\underline H$ and the 
positive roots $E_a$:
\beqa 
&{\rm a)}&\ \ \sigma_K(\underline{k})|K> = K |K>\quad ,\; K > 0\ ,\nonumber\\
&{\rm b)}&\ \ \sigma_K(T(\alpha))|K> = 0\quad {\rm if}\ 
\tr T(\alpha)\underline{k} = 0\ ,\nonumber\\
&{\rm c)}&\ \sigma_K(E_a)|K> = 0\quad {\rm for}\ \forall\ a > 0\ .
\eeqa
A representation of $G$ fulfilling a) and b) always exists: it is
induced from the unitary one-dimensional representation of $H$ given
by a) and b):
\be 
\Sigma_K (e^{i\xi_\alpha T(\alpha)})|K> =
 e^{i\frac{K}{{\rm Tr}\,{\underline k}^2}
{\rm Tr}(\underline{k}\,\xi_\alpha T(\alpha))}|K>\ .\  
\label{dcviii} 
\ee 
Here we have used 
\be
\sigma_K(\xi_\alpha T(\alpha))|K> = \frac{K}{{\rm Tr}\,{\underline
k}^2}{\rm Tr}(\underline{k}\,\xi_\alpha T(\alpha))|K>\ .
\ee
As for a), b) and c) together, it gives the representation of the
group $G_c$ generated by $\underline H$ and $E_a$, $a>0$, induced from 
the representation $\sigma_K(\underline H)$, $\sigma_K(E_a)$, $a>0$. 

Let us fix an orthonormal basis $\{ e_1,e_2,\dots,e_M\}$ in the 
representation space of dimension $M$ (say) of $\Sigma_K(G)$. Choose a
vector $|\zeta)=\sum_{i=1}^M\zeta_ie_i$, $\zeta_i\in{\mathbb C}$, 
$\zeta=(\zeta_1,\zeta_2,\dots ,\zeta_M)$, so that
$(\zeta|\Sigma_K(g)|K>\neq 0$ when $g$ belongs to some open set 
${\cal O}$. Such a $|\zeta)$ exists since $(\zeta|\Sigma_K(g)|K>=1$ for 
$|\zeta)=\Sigma_K(g)|K>$. Further choose ${\cal O}$ so that it is
invariant under $H$-action. That is always possible since for
$(\zeta|\Sigma_K(g)|K>$ changes only by a phase under this action by
(\ref{dcviii}).     

Now the function $\omega_\zeta$ defined by 
\be \omega_\zeta(g) = <\zeta|\Sigma_K(g)|K>   \ ,\quad g\in {\cal O} 
\label{dcix} 
\ee 
has the properties
\beqa
 (X_a\omega_\zeta)(g) &=&  (X_{-a}{\bar\omega}_\zeta)(g) = 0\ \ 
{\rm for}\ \forall a > 0\ ,\nonumber\\
\frac{1}{\omega_\zeta(g)}(X_\alpha\omega)_\zeta(g)&=&
-\frac{1}{{\bar\omega}_\zeta (g)}(X_\alpha{\bar\omega}_\zeta)(g)\ =\
i\frac{K}{{\rm Tr}\,{\underline k}^2}\,\tr(\underline{k}\,T(\alpha))\ . 
\label{dcxii} 
\eeqa
Here the bar denotes complex conjugation. The first line is a 
direct consequence of the fact that $|K>$ is the highest weight
vector, the second line follows from the relation $\omega_\zeta(gh)=
\omega_\zeta(g)\Sigma_K(h)$ valid for any $g\in G$, $h\in H$, with the phase 
factor $\Sigma_K (h)$ given by (\ref{dcviii}).

If $g\in{\cal O}$, the K\"{a}hler potential is given by the formula 
\be 
\Phi_\zeta\ =\ -\frac{{\rm Tr}\,{\underline k}^2}{2K}\, \log\, 
\omega_\zeta(g)\,{\bar\omega}_\zeta(g)\ . 
\label{dcx} 
\ee 
$\Phi_\zeta$ is a function on ${\cal O}/H \subseteq G/H$, since in 
the product 
\[
\omega_\zeta(gh){\bar\omega}_\zeta(gh) = \Sigma_K(h)
\omega_\zeta(g){\bar\Sigma}_K(h){\bar\omega}_\zeta (g) 
\] 
the phase factors $\Sigma_K(h)$ and ${\bar\Sigma}_K(h)$ cancel. The 
K\"{a}hler potential is closely related to the one-form $\Theta$ 
introduced in (\ref{dxcviif}). Thus the exterior derivative $d$ on $G$ 
can be written as $d=d_++d_-+d_0$, where
\be 
d_{\pm}f(g) = (X_{\pm |a|}f)(g)\theta^{\pm |a|}\ ,\ d_0f(g) = (X_\alpha
f)(g)\theta^\alpha\ .
\label{dcii}\ee 
Now, using (\ref{dcxii}) one obtains 
\beqa
i(d_+-d_-)\Phi_\zeta &=& i(X_{|a|}\Phi_\zeta)\,
\theta^{|a|}-i(X_{-|a|}\Phi_\zeta)\, \theta^{-|a|}\nonumber\\ 
&=& -i\frac{{\rm Tr}\,{\underline
k}^2}{2K{\bar\omega}_\zeta}(X_{|a|}{\bar\omega}_\zeta)\,\theta^{|a|}
+i\frac{{\rm Tr}\,{\underline k}^2}{2K\omega_\zeta}
(X_{-|a|}\omega_\zeta)\,\theta^{-|a|} \nonumber\\
&=& i\frac{{\rm Tr}\,{\underline
k}^2}{2K}\,d\log\frac{\omega_\zeta}{{\bar\omega}_\zeta}
+\tr({\underline k}\,T(\alpha))\theta^\alpha=i\frac{{\rm
Tr}\,{\underline k}^2}{2K}\,d\log
\frac{\omega_\zeta}{{\bar\omega}_\zeta}\, +\,\Theta
\label{7.24}
\eeqa 
It follows that 
\be 
d\,i(d_+-d_-)\,\Phi_\zeta\ =\ d\,\Theta\ =\ \Omega\ .
 \label{dcxi} 
\ee
The left hand side of (\ref{dcxi}) can be evaluated using the first
line of (\ref{7.24}) and (\ref{dxcviie}). Calculating its values on  
$iX_a\otimes X_{-b}$ for $0< \lambda_a\leq\lambda_b$ and $0< -\lambda_a 
\leq -\lambda_b$, we get (\ref{6.16}). For this calculation, it
is also important that $c_{abc}=0$ if $\lambda_a+\lambda_b\neq\lambda_c$ .

In another open set ${\cal O}'\subset G$, we may have to work with the 
K\"{a}hler potential $\Phi_\eta$. Then if ${\cal O}\cap{\cal O}'
\neq\emptyset$, the two potentials on ${\cal O}\cap{\cal O}'$ are 
related by
\be
\Phi_\eta\ =\ \Phi_\zeta\, -\,\frac{{\rm Tr}\,{\underline k}^2}{2K}\,
\log\frac{\omega_\eta}{\omega_\zeta}\, +\, 
\frac{{\rm Tr}\,{\underline k}^2}{2K}\, 
\log\frac{{\bar\omega}_\eta}{{\bar\omega}_\zeta}\ .
\ee
The mapping $\Phi_\zeta$ to $\Phi_\eta$ is often called a gauge 
transformation.

\bigskip
{\bf Acknowledgements}

We are part of a collaboration on fuzzy physics and noncommutative
geometry which involves several physicists. The work reported here has
benefited greatly by discussions with the members of this
collaboration. We thank to Franciscus J. Vanhecke for bringing reference 
\cite{cahen} our attention. Our work was supported by D.O.E. and N.S.F
under contract numbers DE-FG02-85ER40231 and INT-9908763, by I.N.F.N,
Italy, and VEGA project 1/7069/20, Slovakia.  

\vfill\break
\bibliographystyle{unsrt}

\end{document}